\documentclass[sigconf,natbib=true,screen,anonymous=false,nonacm]{acmart}

\usepackage{pifont} 

\usepackage{xcolor}  

\usepackage{algorithm}
\usepackage{algorithmicx}
\usepackage{algpseudocode}
\usepackage{multirow}

\AtBeginDocument{%
  \providecommand\BibTeX{{%
    \normalfont B\kern-0.5em{\scshape i\kern-0.25em b}\kern-0.8em\TeX}}}

\setcopyright{acmlicensed}
\copyrightyear{2023}
\acmYear{2023}
\acmDOI{XXXXXXX.XXXXXXX}

\acmConference[Conference acronym 'XX]{Make sure to enter the correct
  conference title from your rights confirmation emai}{June 03--05,
  2018}{Woodstock, NY}
\acmISBN{978-1-4503-XXXX-X/18/06}

\usepackage{booktabs}
\usepackage{xcolor}

\newcommand{\narenc}[1]{\textcolor{red}{Naren writes: #1}}

\title{QuOTE: Question-Oriented Text Embeddings}

\author{Andrew Neeser}
\email{aneeser24@vt.edu}
\affiliation{
    \institution{Virginia Tech}
    \country{USA}
}

\author{Kaylen Latimer}
\email{latime.kaylen22@svvsd.org}
\affiliation{
    \institution{St. Vrain Valley Schools}
    \country{USA}
}

\author{Aadyant Khatri}
\email{aadyant@vt.edu}
\affiliation{
    \institution{Virginia Tech}
    \country{USA}
}

\author{Chris Latimer}
\email{chris.latimer@vectorize.io }
\affiliation{
    \institution{Vectorize.io}
    \country{USA}
}

\author{Naren Ramakrishnan}
\email{naren@vt.edu}
\affiliation{
    \institution{Virginia Tech}
    \country{USA}
}

\begin{document}


\begin{abstract}
We present QuOTE (Question-Oriented Text Embeddings), a novel enhancement to retrieval-augmented generation (RAG) systems, aimed at improving document representation for accurate and nuanced retrieval. Unlike traditional RAG pipelines, which rely on embedding raw text chunks, QuOTE augments chunks with hypothetical questions that the chunk can potentially answer, enriching the representation space. This better aligns document embeddings with user query semantics, and helps address issues such as ambiguity and context-dependent relevance. Through extensive experiments across diverse benchmarks, we demonstrate that QuOTE significantly enhances retrieval accuracy, including in multi-hop question-answering tasks. Our findings highlight the versatility of question generation as a fundamental indexing strategy, opening new avenues for integrating question generation into retrieval-based AI pipelines.

\end{abstract}


\settopmatter{printfolios=true}

\maketitle

\section{Introduction}

Retrieval-augmented generation (RAG~\cite{rag1,rag2,rag3})
serves as a significant contribution to the deployment and acceptance of LLMs in practice.
Given a user's prompt, RAG retrieves relevant information from a document collection, augments (prefixes) it to the user's prompt, thus helping ensure that any generated content can be accurate, pertinent, and grounded in up-to-date information.
In a typical RAG implementation, at pre-query time, the corpus is broken down into chunks, which are stored as vector embeddings. At query time,  these chunks are searched and used to augment the user's prompt.
Several variants of RAG have been proposed over the years~\cite{rag-variant1, rag-variant2, rag-variant3} to address specific use cases and challenges. 

RAG has helped reinforce the criticality of information retrieval (IR) as a vital component of modern NLP and AI pipelines. Despite this resurgence, much of the focus has been on enhancing the G (generation) component, often leaving advancements in the R (retrieval) aspect comparatively underexplored. Recently, some notable efforts have emerged to address this imbalance.

For example, Anthropic introduced contextual retrieval~\cite{anthropic_contextual_retrieval}
where each
chunk is augmented with additional context before embedding; this approach is claimed to reduce incorrect chunk retrieval rates by up to 67\%.
Similarly, recent works have explored prompt caching~\cite{prompt-caching-RAG}, a strategy to reuse previously retrieved or generated results to optimize latency and computation costs in iterative or repetitive query scenarios.

\begin{figure}[ht]
    \centering
    \includegraphics[width=1\columnwidth]{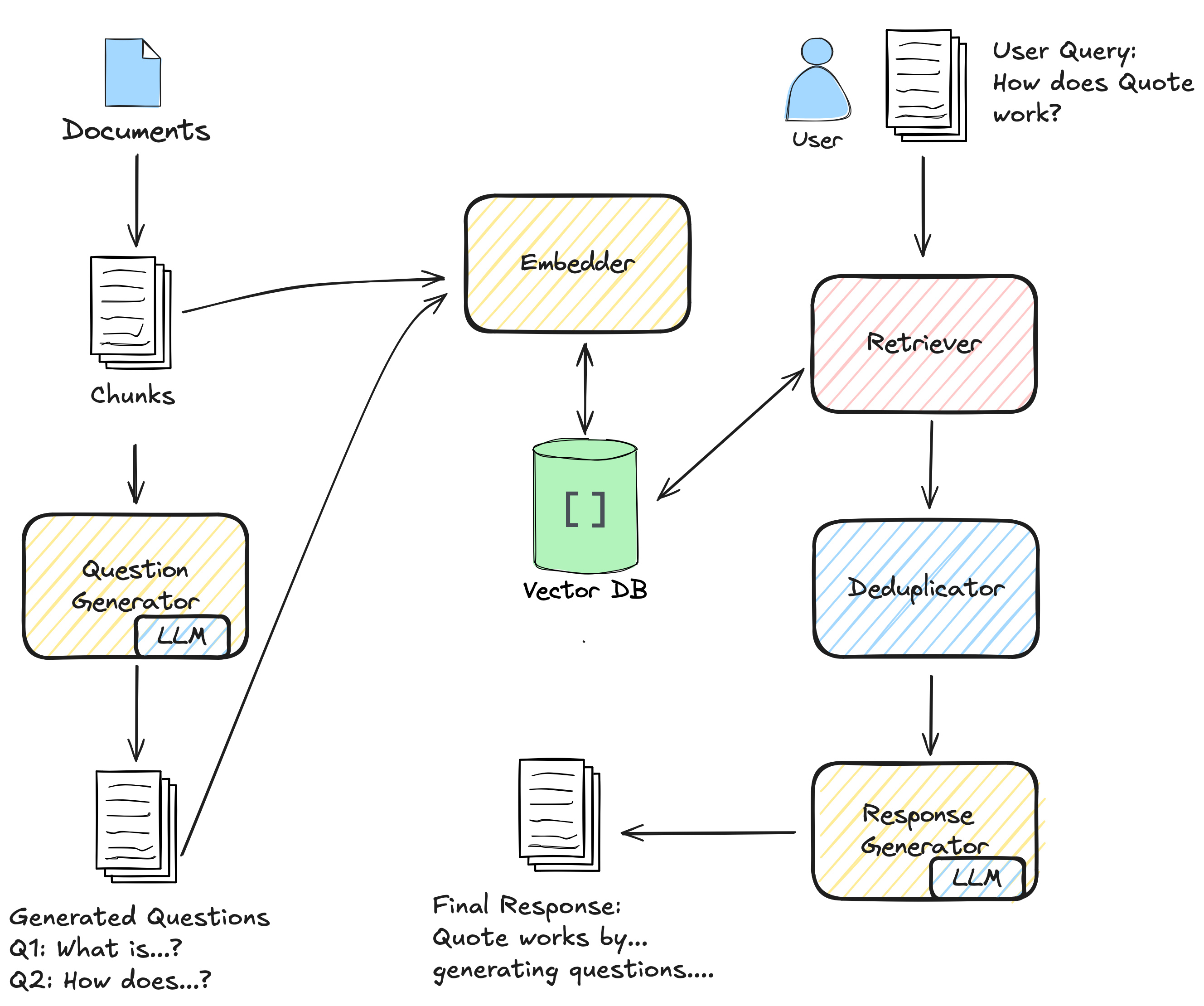}
    \caption{Overview of QuOTE. Documents are split into chunks and processed by a question generator (LLM) to create relevant questions. Chunks along with the questions they purport to answer are embedded in a vector database. At query time, a retriever and deduplicator processes user queries to generate final responses.}
    \label{fig:quote_overview}
\end{figure}

Our work aligns with this vein of `advancing R for G', particularly focusing on improving the modeling of document chunks as they are embedded. One of our key insights is that documents can often be more effectively represented by the questions they can answer, rather than solely by their direct content. To this end, for each chunk, we propose generating a set of questions that the chunk is likely to answer, embedding these alongside the original content. We refer to such embeddings as Question-Oriented  Text Embeddings (QuOTE). See Fig.~\ref{fig:quote_overview} for how QuOTE works.

This paper makes the following contributions.
\begin{enumerate}
\item We demonstrate that 
the idea of embedding (hypothetical) questions along with text chunks significantly enhances retrieval performance, particularly in scenarios where nuanced understanding of the content is required. This idea holds promise beyond RAG by opening up the possibility of question generation as a fundamental indexing strategy.
\item 
We prioritize retrieval performance 
rather than
generation quality in our evaluation, and
conduct an exhaustive empirical analysis of QuOTE with multiple language models, several key datasets, a range of query workloads, and compare it versus other RAG benchmarks. This gives insight into specific regions of the configuration space where QuOTE performs best and future directions of research.
\item Beyond empirical results, we characterize the features of RAG settings 
where (and why) QuOTE works, and how we can anticipate performance improvements prior to embarking on QuOTE-style indexing for given corpora. 
\end{enumerate}

\section{Related Work}
Many studies have
highlighted the impact of key design choices for the success of a RAG implementation~\cite{ragimp1,ragimp2, ragimp3}. 

\subsection{Dense vs Sparse Retrievers}
The debate between dense and sparse retrievers continues into
RAG
research~\cite{sparse-vs-dense1,sparse-vs-dense2}. Dense retrievers, such as those based on vector embeddings, excel at capturing semantic similarity, making them particularly effective for nuanced queries. However, sparse retrievers like BM25 and TF-IDF continue to dominate in scenarios where explicit token matches, such as named entities, acronyms, or abbreviations, are critical to relevance. This distinction has led to hybrid approaches in many RAG systems, which combine dense and sparse retrievers. For example, a typical implementation involves first running a keyword-based sparse retrieval to gather an initial pool of relevant chunks, followed by a dense retrieval to refine the results. 

\subsection{Retrievers vs Rerankers}
Many RAG systems employ a two-step pipeline: a fast retriever selects the top-k candidate chunks, and a reranker, typically a computationally intensive cross-encoder, reorders these candidates for final use. While rerankers generally improve the quality of retrieved results, recent research~\cite{Jacob-Drozdov-drowning-in-documents} cautions against extending reranking to larger candidate sets. Beyond a certain threshold, performance tends to plateau and may even degrade, likely due to noise introduced in larger retrieval pools. These findings underscore the importance of balancing efficiency and effectiveness in the retrieval-reranking pipeline.

\subsection{Exact search vs Approximate Nearest Neighbors (ANN)}
Approximate nearest neighbor (ANN) techniques ~\cite{ann-ir-paper} have become the de facto standard for scalable dense retrieval due to their ability to handle large corpora efficiently. However, exact search methods, while computationally more demanding, offer greater precision in certain use cases, such as high-stakes QA tasks. Several studies~\cite{anncomp1, anncomp2} compare these approaches, highlighting trade-offs in latency, accuracy, and robustness to query variations. For instance, ANN methods may struggle with long-tail queries or datasets containing subtle semantic distinctions.

\subsection{Distractions vs Noise in RAG}
Cuconasu et al.~\cite{cuconasu} study the performance of RAG for QA tasks in the presence of so-called {\it distracting} and {\it noise} documents. Distracting documents are those with high retrieval scores, but that do not contain the answer; noise documents are picked at random from the corpus. The interesting finding from this study was that while distracting documents lead to performance deterioration as expected, noise documents lead to improved performance, presumably due to better reliance on pretrained reasoning. 
However, these findings are somewhat questioned by recent work~\cite{leto2024toward}, which suggests that noise documents can degrade system reliability in certain settings, calling for further investigation.


\subsection{Real vs Hypothetical Embeddings}
Contextual retrieval techniques, such as Anthropic's approach to augmenting chunks with additional information before embedding, have emerged as promising ways to reduce retrieval errors. Similarly, Hypothetical Document Embeddings (HyDE)~\cite{hyde2022} involve generating synthetic text based on the query and embedding it alongside real documents. These methods aim to capture query-specific nuances, resulting in more robust retrieval in open-domain and QA contexts. Our work builds on these approaches by leveraging question-based chunk representations for improved relevance.


\subsection{Supporting Asymmetric QA Tasks}
In many QA scenarios, particularly in customer support and enterprise search, there exists a fundamental asymmetry: user queries are often brief, while answers require detailed, structured information. RAG systems addressing this imbalance have incorporated techniques such as hierarchical retrieval~\cite{raga}, multi-hop reasoning~\cite{mavi2022multihop}, and weighted retrieval pipelines~\cite{ragc} to bridge this gap. Recent efforts in this domain include query-expansion strategies~\cite{ragf} and retrieval conditioning~\cite{ragg} to better align user intent with document granularity.

\subsection{Neural Information Retrieval}
Neural information retrieval methods aim to model complex semantic relationships and contextual relevance more effectively than traditional approaches.
While approaches like ColBERT~\cite{colbert-3} and DPR~\cite{dpr-4} have made significant strides in dense retrieval, they continue to struggle with nuanced information seeking behaviors, involving hierarchical relationships, managing distributed information across multiple documents, and dealing with context-dependent relevance ranking.

\subsection{End-to-End RAG Systems}
Fully integrated, end-to-end RAG systems (e.g., from companies like Vectorize.io) are becoming increasingly popular for tasks requiring seamless interaction between retrieval and generation. Recent work~\cite{10.1145/3626772.3657957} has focused on optimizing these systems for efficiency, scalability, and robustness. End-to-end designs often integrate prompt caching, hybrid retrieval, and adaptive reranking to achieve state-of-the-art performance across diverse NLP tasks.


\section{QuOTE}

\begin{algorithm}[ht]
\caption{Building the QuOTE Index (Pseudocode)}
\label{algo:build-quote-index}
\begin{algorithmic}[1]
\Require Corpus C, LLM, VectorDB, NumQuestions
\Function{BuildQuoteIndex}{C, LLM, VectorDB, NumQuestions}
    \State $\mathcal{P} \gets \text{SplitCorpus}(C)$ \Comment{Split into chunks/passages}
    \ForAll{$p \in \mathcal{P}$}
        \State $Q \gets \text{LLMGenerateQuestions}(p,\text{NumQuestions})$
        \ForAll{$q \in Q$}
            \State $doc \gets q \| p$ \Comment{concatenate or store separately}
            \State \text{VectorDB.Add}(doc, \text{metadata}=\{\text{originalChunk}: p\})
        \EndFor
    \EndFor
\EndFunction
\end{algorithmic}
\end{algorithm}

\begin{algorithm}[ht]
\caption{Querying the QuOTE Index (Pseudocode)}
\label{algo:quote-query}
\begin{algorithmic}[1]
\Require UserQuery $u$, VectorDB, $k$, $M$
\Statex
\Function{QuoteQuery}{$u$, VectorDB, $k$, $M$}
    \State $u_{emb} \gets \Call{Embed}{u}$
    \State $\mathcal{R} \gets \Call{VectorDB.Query}{u_{emb}, top=k \times M}$
    \State uniqueResults $\gets \Call{Deduplicate}{\mathcal{R}}$
    \State finalContexts $\gets \Call{TopK}{uniqueResults, k}$
    \State answer $\gets \Call{LLM}{UserQuery \parallel finalContexts}$
    \State \Return answer
\EndFunction
\end{algorithmic}
\end{algorithm}

QuOTE can be viewed
in the lineage of
query reformulation~\cite{query-formulation-6}
and multi-hop reasoning~\cite{multi-hop-7}, but takes a unique perspective by focusing on question generation as a fundamental indexing strategy. 
As discussed earlier, the naive RAG approach can
fail to capture the \emph{intent} behind user queries, especially when queries are succinct 
(e.g., entity lookups) or require extracting specific details from a chunk. 
In QuOTE we transform
each chunk of text into \emph{multiple}
(question + chunk) representations
capturing a range of opportunities for retrieval.
Note that each generated question (plus chunk) is then stored as a separate ``document'' or embedding in the vector database. 

See Algorithm~\ref{algo:build-quote-index} for pseudocode to illustrate how QuOTE builds an index, 
and 
Algorithm~\ref{algo:quote-query}
for how it is queried.
We next describe key stages of the pipeline (see Fig. \ref{fig:quote_overview}):

\subsection{Question Generation at Pre-Query Time}
We split the corpus into smaller passages (or chunks). For each
chunk, we prompt an LLM to generate a set of questions that the chunk can answer. While question generation is a well studied topic in NLP~\cite{heilman_smith_2009, qgen2, heilman_smith_2010}. 
The quality and diversity of generated questions play a significant role in QuOTE’s effectiveness. We use an LLM with
prompt engineering (see Section~\ref{subsec:prompt-effects}) to create a representative set of questions with specificity and coverage. By creating multiple question-based embeddings for each chunk, QuOTE better captures diverse user queries that reference the same text in different ways. 
If a user’s query is similar (semantically) to one of the chunk-generated questions, that chunk becomes more likely to rank highly, 
leading to more accurate retrieval.

\subsection{Embedding}
 Instead of storing just the original chunk embedding, \emph{we store each generated question} (along with the original chunk) in the vector database. 
 In Section~\ref{sec:quote-orthogonality} we demonstrate that the performance
 of QuOTE is agnostic to the
 choice of embedding model.

 \subsection{Retrieval and Deduplication at Query Time}
 During query time, multiple retrieved
``documents'' often reference the same underlying chunk. Hence,
a deduplication step is necessary to ensure we select the top-k distinct chunks, avoiding wasted slots.
To this purpose we `over-retrieve'
top-$k \times M$ results (for some value of $M$) from the question-based embeddings. 
(Note that this de-duplication step is unique to the QuOTE pipeline and is not a feature in classical RAG pipelines.)

\section{Datasets and Metrics}

\begin{table*}[t]
\centering
\caption{Overview of candidate datasets for RAG evaluation. \textbf{SQuAD}, \textbf{MultiHop-RAG}, and \textbf{Natural Questions} are included to help evaluate retrieval performance.
Other datasets (\textbf{ELI5}, \textbf{HotpotQA}, \textbf{Frames}, \textbf{TriviaQA}) were considered but are excluded either because they are geared toward assessing generation quality or for other reasons described.}
\label{tab:dataset-comparison}
\renewcommand{\arraystretch}{1.15}
\small
\begin{tabular}{p{2.2cm} p{7.0cm} p{1.8cm} p{5.5cm}}
\toprule
\textbf{Dataset} & \textbf{Description / Focus} & 
\textbf{Included or Excluded?} &
\textbf{Rationale} \\
\midrule

\textbf{SQuAD}~\cite{squad1} 
& 
Over 100k crowd-sourced QA pairs from Wikipedia articles (single-hop). 
Requires precise extractive answers (spans).
& 
\textcolor{green}{\Huge \checkmark}
& Classic reading comprehension benchmark with moderate passage lengths. Ideal for testing single-hop retrieval. \\

\textbf{MultiHop-RAG}~\cite{multihoprag-paper}
& 
2{,}555 queries requiring retrieval from multiple documents. 
Each question links to a list of fact sentences across different sources, ensuring genuine multi-hop reasoning.
& 
\textcolor{green}{\Huge \checkmark}
&
Explicitly tests multi-document retrieval and cross-sentence reasoning. 
Straightforward corpus chunking (4-sentence blocks) with traceable ground truth. \\

\textbf{Natural Questions (NQ)}~\cite{nq}
& 
Real Google Search queries mapped to Wikipedia, featuring both short and long answers. 
Demands chunking of lengthy documents.
& 
\textcolor{green}{\Huge \checkmark}
&
Reflects real user queries. Large scale and diverse. 
Useful for evaluating open-domain QA in a practical setting. \\

\textbf{ELI5}~\cite{eli5}
& 
A long-form QA dataset (272k questions from the “Explain Like I’m Five” subreddit). 
Open-ended questions requiring paragraph-length answers.
& 
\textcolor{red}{\Huge \ding{55}}
&
Official data no longer accessible due to Reddit API changes. 
No well-defined short-answer ground truth chunks; original passages are difficult to re-obtain. \\

\textbf{HotpotQA}~\cite{hotpotqa}
& 
113k Wikipedia-based QA pairs designed for multi-hop reasoning. 
Includes supporting sentence-level “facts.”
& 
\textcolor{red}{\Huge \ding{55}}
&
Context is provided in disjointed sentence lists, complicating chunk-based retrieval. 
MultiHop-RAG’s more document-centric structure provides a natural benchmark. \\

\textbf{Frames}~\cite{frame} 
& 
824 multi-hop questions requiring 2–15 Wikipedia articles. 
Complex domain with no provided corpus (web scraping needed).
& 
\textcolor{red}{\Huge \ding{55}}
&
Highly specialized queries with small question count (824), 
and no pre-fetched Wikipedia corpus. Overly complex to integrate. \\

\textbf{TriviaQA}~\cite{triviaqa}
& 
650k question-answer-evidence triples from various sources, often requiring multi-document reasoning. 
Focuses on “trivia-style” queries.
& 
\textcolor{red}{\Huge \ding{55}}
&
Not all questions have explicit matching documents. 
Already have coverage of multi-hop via MultiHop-RAG. 
Would require extensive chunking / ground-truth alignment. \\

\bottomrule
\end{tabular}
\label{dataset-table}
\end{table*}

While there exist a variety of datasets for RAG evaluation (see Table~\ref{dataset-table}) not all are geared toward evaluating retrieval performance as distinct from generation, which is our focus here. For instance, QA datasets where we are evaluated against the quality of the generated answer, or where the original ground truth chunks are not available, do not support assessing the performance of QuOTE in helping improve retrieval of relevant chunks. Accordingly, we focus on 
three benchmark datasets commonly used for question answering: 
\emph{Natural Questions} (NQ)~\cite{nq}, \emph{SQuAD}~\cite{squad1,squad2}, and \emph{MultiHop-RAG}~\cite{multihoprag-paper}. 
These datasets vary in complexity, domain coverage, and the style of questions, providing a broad platform to test the retrieval capabilities of our approach.

\subsection{Natural Questions (NQ)}
\label{subsec:nq}
The Natural Questions (NQ) dataset~\cite{nq} is a large-scale benchmark,
with questions directly sourced
from real user queries and answers keyed to Wikipedia articles. The dataset is split into 
approximately 307k training examples and roughly 7.8k each in the development and test sets. 
For each query, the dataset provides the relevant passages (long answer) and the precise phrases or entities (short answer) 
where the answer resides.

One non-trivial issue pertains to {\it multiple, highly similar passages 
in the same article}.
For example, consider passages about the song ``'Heroes''' by David Bowie from the Wikipedia article titled ``Heroes (David Bowie song)''. This article has two passages that
are nearly identical, differing only in minor wording (e.g., “in the UK” vs. “in the United Kingdom”). These slight variations do not change the factual content but result in multiple, nearly duplicate contexts.

Such minor differences unnecessarily fragment the dataset into multiple contexts, each labeled as distinct. This discrepancy complicates retrieval-based evaluations because systems are penalized if they return an almost-correct chunk that differs by only a few words from the one labeled as ground-truth.

To address this issue, we merge highly similar chunks based on a text-similarity threshold, combining their respective questions into a single context group. This merging strategy reduces noise and ensures that semantically equivalent passages (or chunks) are treated as one, allowing retrieval mechanisms to focus on true distinctions in content rather than trivial rephrasings.

\subsection{SQuAD} The Stanford Question Answering Dataset (SQuAD)~\cite{squad1} is widely recognized as a benchmark for reading comprehension 
and extractive QA. 
Each question is associated with an exact answer span in the corresponding article, ideal for our extractive evaluation purposes.

\subsection{MultiHop-RAG} \emph{MultiHop-RAG}~\cite{multihoprag-paper} is specifically designed to test multi-hop question answering. 
Unlike SQuAD and NQ, which pair each question with a single relevant paragraph or article, 
MultiHop-RAG associates multiple ground-truth documents with each query. 
For instance, a query such as:
\emph{``Which company is being scrutinized by multiple news outlets for anticompetitive practices 
and is also suspected of foul play by individuals in other reports?''}
will require cross-referencing two or more articles to gather the necessary evidence. 

\subsection{Evaluation Metrics}
\label{subsec:nq-metrics}
For \emph{Natural Questions (NQ)}
and \emph{SQuAD},
each query typically has a \emph{single} correct Wikipedia article and a specific paragraph in that article as ground truth. 
We use the following metrics aimed at capturing whether QuOTE can precisely isolate the article along with the correct answer span.

\begin{itemize}
    \item \textbf{Context Accuracy (C@k):} The fraction of queries for which the correct \emph{paragraph-level} context 
    is retrieved within the top-$k$ results. If a system retrieves the exact paragraph containing the short
    answer at any rank $\le k$, we consider it a successful retrieval.

    \item \textbf{Title Accuracy (T@k):} The fraction of queries for which the correct \emph{article-level} title 
    is found among the top-$k$ results. This is a coarser (i.e., easier) measure compared to paragraph-level context accuracy 
    but still offers insight into whether the system can identify the right document (for instance, the correct Wikipedia page).
\end{itemize}

\noindent
\emph{MultiHop-RAG} queries can reference \emph{multiple} relevant documents. Consequently, we employ:

\begin{itemize}
    \item \textbf{Full Match Accuracy (Full@k):} All evidence pieces required by the query must be found 
    within the top-$k$ retrieved results. If even one piece of critical evidence is missing, 
    the query is marked as a failure under this measure.

    \item \textbf{Partial Match Accuracy (Part@k):} Because missing one or more documents can still lead to a partially correct answer, 
    we measure the \emph{percentage of required evidence} found in the top-$k$ results. This measure 
    highlights how retrieval errors degrade performance. For instance, a system might retrieve 2 of the 3 
    needed documents (66.7\% partial match), which can be useful for partial downstream reasoning but might 
    not yield the fully correct answer.
\end{itemize}

This two-level evaluation (full vs.\ partial) captures the difficulty of multi-hop retrieval 
where multiple documents must be combined to arrive at a final answer.
\section{Evaluation}

We conduct a comprehensive evaluation
to answer the below questions:
\begin{enumerate}
\item (Section~\ref{subsec:prompt-effects}) Is QuOTE able to automatically generate questions that improve the performance of retrieval-augmented generation?
\item 
(Section~\ref{sec:quote-orthogonality})
How sensitive is QuOTE performance to the choice of embedding model?
\item (Section~\ref{subsec:num-questions})
How many questions must be generated for QuOTE to be effective?
\item (Section~\ref{subsec:compare-hyde}) How does QuOTE compare to HyDE, the state-of-the-art approach to query enrichment? 
\item (Section~\ref{subsec:dedup-effect}) How negligible or significant is QuOTE's deduplication overhead?
\item (Section~\ref{subsec:cheaper-models}) Because QuoTE uses an LLM for question generation as well as for answer generation, can we employ a cheaper model for question generation and does this significantly affect performance?
\item (Section~\ref{subsec:context-vs-accuracy}) Can we characterize the properties of contexts for which QuOTE has selective superiority?
\end{enumerate}

\subsection{Effect of Different Prompts to Generate Questions}
\label{subsec:prompt-effects}

\begin{table*}[t]
\centering
\caption{Prompt templates for 
Natural Questions (NQ), SQuAD, and
MultiHop-RAG.}
\label{tab:all-prompts}
\renewcommand{\arraystretch}{1.1}
\begin{tabular}{p{3.2cm} p{13.0cm}}
\toprule
\textbf{Dataset + Prompt} & \textbf{Prompt Text (Single String)} \\
\midrule

\textbf{NQ and SQuAD (Basic)}
&
\footnotesize
\texttt{"Generate numerous questions to properly capture all the important parts of the text. Separate each question-answer pair by a new line only; do not use bullets. Format each question-answer pair on a single line as 'Question? Answer' without any additional separators or spaces around the question mark. Text:\textbackslash\{chunk\_text\textbackslash\}"} \\[1em]

\textbf{NQ and SQuAD (Complex)}
&
\footnotesize
\texttt{"Read the following text and generate numerous factual question-answer pairs designed to resemble authentic user search queries and natural language variations. Each question should accurately and semantically capture important aspects of the text, with varying lengths and complexities that mirror real-world search patterns. Include both shorter, keyword-focused questions such as 'who founded Tesla Motors' and longer, natural style questions like 'when did Elon Musk first start Tesla company'. Incorporate 'how' and 'why' questions to reflect genuine user curiosity. Avoid using phrases like 'according to the text' and abstain from pronouns by specifying names or entities. Ensure questions are not overly formal or artificial, maintaining a natural query style. Immediately follow each question with its precise answer on the same line, formatted as 'Question? Answer', without any additional formatting or commentary. Each pair should be on its own line. Text:\textbackslash\{chunk\_text\textbackslash\}"} \\[1em]
\textbf{MultiHop-RAG (Basic)} 
& 
\footnotesize
\texttt{"Generate enough multi-hop questions along with their answers to properly capture all the important parts of the text. These questions should require integrating multiple pieces of information to answer. Separate each question-answer pair by a new line only; do not use bullets. Format each question-answer pair on a single line as 'Question? Answer' without any additional separators or spaces around the question mark. Text:\textbackslash\{chunk\_text\textbackslash\}"} \\[1em]
\textbf{MultiHop-RAG (Complex)} 
& 
\footnotesize
\texttt{"Read the following text and generate complex, multi-hop questions that require integrating multiple pieces of information from the text to answer. The questions should involve reasoning and synthesis, referring to different parts or aspects of the text. Do not use phrases like 'according to the text', 'mentioned in the text', or 'in the text'. All questions should be one sentence long. Never use pronouns in questions; instead, use the actual names or entities. Format each question on a single new line as Question? Answer without any additional separators or spaces around the question mark. Text:\textbackslash\{chunk\_text\textbackslash\}"} \\[1em]
\bottomrule
\end{tabular}
\end{table*}

\begin{table*}[t]
\centering
\small
\caption{Key retrieval results across Natural Questions (NQ), SQuAD, and MultiHop-RAG for 
\textbf{Naive}, \textbf{Basic}, and \textbf{Complex} prompting strategies. 
For NQ and SQuAD, we report \textbf{Top-1} (\(C@1\)) and \textbf{Top-5} (\(C@5\)) Context Accuracy, 
along with \textbf{Top-1} (\(T@1\)) and \textbf{Top-5} (\(T@5\)) Title Accuracy. 
For MultiHop-RAG, we present \textbf{Full Match} at \(k=5\) and \(k=20\) (\(\text{Full@}5\), \(\text{Full@}20\)), 
and \textbf{Partial Match} (\(\text{Part@}5\), \(\text{Part@}20\)). 
Bolded entries denote the best performance for each metric.}
\label{tab:combined-results}
\renewcommand{\arraystretch}{1.1}
\begin{tabular}{l|cccc|cccc|cccc}
\toprule
\multirow{2}{*}{\textbf{Method}} 
& \multicolumn{4}{c|}{\textbf{Natural Questions (NQ)}} 
& \multicolumn{4}{c|}{\textbf{SQuAD}} 
& \multicolumn{4}{c}{\textbf{MultiHop-RAG}} \\
\cmidrule(lr){2-5}\cmidrule(lr){6-9}\cmidrule(lr){10-13}
& \textbf{C@1} & \textbf{C@5} & \textbf{T@1} & \textbf{T@5} 
& \textbf{C@1} & \textbf{C@5} & \textbf{T@1} & \textbf{T@5}
& \textbf{Full@5} & \textbf{Full@20} & \textbf{Part@5} & \textbf{Part@20} \\
\midrule
\textbf{Naive} 
& 32.92\% & 89.23\% & \textbf{99.85\%} & \textbf{100.00\%}
& 82.58\% & 96.00\% & 98.70\% & 99.13\%
& 8.00\%  & 21.50\% & 29.6\%  & 50.0\%  \\
\textbf{Basic} 
& 34.77\% & \textbf{92.46\%} & \textbf{99.85\%} & \textbf{100.00\%}
& 85.82\% & 98.16\% & 99.46\% & \textbf{99.89\%}
& 3.00\%  & 16.50\% & 27.7\%  & 47.5\%  \\
\textbf{Complex} 
& \textbf{38.00\%} & 92.15\% & 99.69\% & \textbf{100.00\%}
& \textbf{90.26\%} & \textbf{98.81\%} & \textbf{99.68\%} & 99.78\%
& \textbf{8.50\%} & \textbf{26.50\%} & \textbf{31.7\%} & \textbf{54.9\%} \\
\bottomrule
\end{tabular}
\end{table*}

A central consideration for QuOTE-style indexing is how the prompt itself influences the \emph{quality} of generated questions. We compare two main prompt templates:
\begin{itemize}
    \item \textbf{Basic Prompt:} Instructs the model to “Generate enough questions to properly capture all the important parts of the text”. The questions are short, direct, and do not include advanced reasoning cues.
    \item \textbf{Complex Prompt:} Adds instructions for more detailed or multi-hop reasoning. In MultiHop-RAG, for example, the complex prompt explicitly requests multi-hop questions referencing multiple pieces of information. In SQuAD or NQ, it encourages short factual queries without referencing the text directly, thereby aiming for more robust coverage of the chunk’s content.
\end{itemize}
\noindent
We compare the performance of both these prompts with a naive RAG implementation.
As
Table~\ref{tab:combined-results}
shows,
the Complex Prompt achieves the highest
Top-1 Context Accuracy overall.
Title Accuracy metrics remain near-perfect across all methods beyond Top-1, indicating that differences among prompts are most pronounced at the paragraph selection level. These observations suggest that more advanced prompting yields modest but meaningful improvements in precisely identifying relevant
questions for specific
passages.

\subsection{QuOTE Performance vis-a-vis Embedding Model}
\label{sec:quote-orthogonality}

Table~\ref{tab:embeddings-comparison} compares \textbf{Naive} vs.\ \textbf{QuOTE} 
 modes on three datasets---\textbf{SQuAD} (single-hop), \textbf{NQ} (single-hop), 
and \textbf{MultiHop-RAG} (multi-hop) across a range of datasets. We report Top-$k$ context/title accuracy for SQuAD and NQ, 
and full/partial match for MultiHop. Despite large differences in baseline quality
(e.g., \texttt{jinaai} vs.\ \texttt{WhereIsAI} vs.\ \texttt{Alibaba}),
note that QuOTE generally improves 
retrieval metrics (especially Top-1 Context Accuracy or Full@20) \emph{regardless of the underlying embedding model}.
QuOTE often raises Top-1 Context Accuracy by 5--17 points on SQuAD and 1--3 points on NQ, and can improve Full@20 by up to several points in MultiHop-RAG.
MultiHop-RAG remains challenging, as even large gains may yield relatively modest absolute numbers (e.g., 9\% or 10\% full match at \(k=5\)). However, QuOTE still outperforms or closely matches a naive approach across all embedding models.


\begin{table*}[t]
\centering
\small
\caption{Performance of \textbf{Naive} vs.\ \textbf{QuOTE} modes on SQuAD, NQ, and MultiHop-RAG, 
across five embedding models. Per-row bolded entries denote the better value for that metric. 
}
\label{tab:embeddings-comparison}
\renewcommand{\arraystretch}{1.05}
\begin{tabular}{ll|cc|cc|cccc}
\toprule
\multirow{2}{*}{\textbf{Embedding Model}} 
& \multirow{2}{*}{\textbf{Approach}} 
& \multicolumn{2}{c|}{\textbf{SQuAD}} 
& \multicolumn{2}{c|}{\textbf{NQ}} 
& \multicolumn{4}{c}{\textbf{MultiHop-RAG}} \\
\cmidrule(lr){3-4}\cmidrule(lr){5-6}\cmidrule(lr){7-10}
& 
& \textbf{C@1} & \textbf{C@20}
& \textbf{C@1} & \textbf{C@20}
& \textbf{Full@5} & \textbf{Full@20} & \textbf{Part@5} & \textbf{Part@20} \\
\midrule
\multirow{2}{*}{\texttt{Alibaba-NLP/gte-base-en-v1.5}} 
& Naive 
  & 59.94 & 96.35
  & 54.25 & 91.59
  & 6.50 & 25.50 & 29.2 & 55.7 \\
& QuOTE 
  & \textbf{77.16} & \textbf{97.68}
  & \textbf{57.18} & \textbf{92.72}
  & \textbf{9.00} & \textbf{27.50} & \textbf{32.0} & \textbf{56.9} \\
\midrule
\multirow{2}{*}{\texttt{text-embedding-3-small}} 
& Naive 
  & 68.59 & \textbf{97.58}
  & 51.15 & \textbf{92.59}
  & 7.00 & \textbf{32.50} & 33.6 & \textbf{62.0} \\
& QuOTE   
  & \textbf{79.07} & 96.69
  & \textbf{51.57} & 90.16
  & 7.00 & 27.50 & \textbf{33.7} & 56.0 \\
\midrule
\multirow{2}{*}{\texttt{WhereIsAI/UAE-Large-V1}}
& Naive
  & 69.49 & 96.98
  & 56.84 & 94.14
  & 4.00 & 23.00 & \textbf{31.8} & \textbf{57.9} \\
& QuOTE
  & \textbf{80.64} & \textbf{97.51}
  & \textbf{58.35} & \textbf{94.31}
  & \textbf{9.00} & \textbf{27.00} & 31.4 & 54.9 \\
\midrule
\multirow{2}{*}{\texttt{jinaai/jina-embeddings-v3}}
& Naive
  & 65.45 & 94.45
  & 53.87 & \textbf{93.76}
  & 4.50 & \textbf{25.00} & 26.0 & \textbf{53.0} \\
& QuOTE
  & \textbf{76.97} & \textbf{96.94}
  & \textbf{55.17} & 93.72
  & \textbf{6.50} & 24.00 & \textbf{29.9} & 51.6 \\
\midrule
\multirow{2}{*}{\texttt{sentence-transformers/all-MiniLM-L6-v2}}
& Naive
  & 65.53 & 95.03
  & 50.77 & 87.23
  & 5.00 & \textbf{20.50} & 26.8 & \textbf{51.2} \\
& QuOTE
  & \textbf{72.35} & \textbf{97.75}
  & \textbf{51.19} & \textbf{88.57}
  & \textbf{5.50} & 20.00 & \textbf{26.9} & 47.0 \\
\bottomrule
\end{tabular}
\end{table*}

\subsection{Effect of Number of Questions}
\label{subsec:num-questions}

One key factor in \emph{question-oriented} retrieval is deciding how many questions an LLM should generate for each chunk of text. 
Generating too few may overlook critical details, while generating too many can introduce redundancy or noise. 
We therefore tested multiple settings across our three datasets (\emph{Natural Questions}, \emph{SQuAD}, and \emph{MultiHop-RAG}), 
varying the number of questions (1, 5, 10, 15, 20, 30) and also including an `LLM decides' setting.
In each case, we measure how \textbf{Context Accuracy} and \textbf{Title Accuracy} changes, or in the case of MultiHop-RAG, 
how \textbf{Full Match} and \textbf{Partial Match} scores are affected.

To systematically investigate the effect of varying the number of generated questions, we parameterize our LLM prompt to either generate:
\begin{itemize}
    \item \textbf{Fixed \# Questions:} If a desired quantity \textit{num\_questions} is provided, the prompt includes a directive such as:

\begin{verbatim}
"Generate exactly {num_questions} questions 
to properly capture all the important parts
of the text."
\end{verbatim}

    \item An \textbf{LLM Decides \# Questions}: Here, the LLM is simply instructed to:

\begin{verbatim}
"Generate enough questions to properly capture
all the important parts of the text."
\end{verbatim}
\end{itemize}



\paragraph{SQuAD}
Table~\ref{tab:questions-comparison} shows that as the number of generated questions per chunk increases from 5 
to around 10 or 20, Top-1 Context Accuracy rises from about 73\% to as high as 76\%, and \emph{Top-5} surpasses 97\% in most settings. 
\emph{Title Accuracy} also remains consistently high, crossing 99\% even at Top-1 for 10+ questions. 
Interestingly, letting the LLM decide how many questions to generate (``LLM Decides'') yields a strong Top-1 Context Accuracy of 76.17\% 
and Top-1 Title Accuracy of 99.30\%.

\begin{itemize}
    \item \textbf{Naive vs. 10 questions.} A naive approach (66.60\% Top-1 Context) significantly lags behind 
    generating 10 questions (74.91\% Top-1), showing that question augmentation dramatically helps correct chunk retrieval.
    \item \textbf{Diminishing returns.} Beyond 10–15 questions, the gains in Top-1 Context Accuracy plateau 
    around 74–76\%. For instance, 20 questions achieve 76.66\%, comparable to 10 questions at 74.91\%.
\end{itemize}

\paragraph{Natural Questions (NQ)}

\begin{table*}[t]
\centering
\caption{Performance comparison across different numbers of generated questions on SQuAD, NQ, and MultiHop-RAG datasets. Results show Context and Title Accuracy at different k values for SQuAD and NQ, and Full/Partial Match for MultiHop-RAG. Bolded entries denote the best performance per metric.}
\label{tab:questions-comparison}
\renewcommand{\arraystretch}{1.05}
\begin{tabular}{l|cccc|cccc|cccc}
\toprule
\multirow{2}{*}{\textbf{Questions}} 
& \multicolumn{4}{c|}{\textbf{SQuAD}} 
& \multicolumn{4}{c|}{\textbf{NQ}} 
& \multicolumn{4}{c}{\textbf{MultiHop-RAG}} \\
\cmidrule(lr){2-5}\cmidrule(lr){6-9}\cmidrule(lr){10-13}
& \textbf{C@1} & \textbf{C@5} & \textbf{T@1} & \textbf{T@5}
& \textbf{C@1} & \textbf{C@5} & \textbf{T@1} & \textbf{T@5}
& \textbf{Full@5} & \textbf{Full@20} & \textbf{Part@5} & \textbf{Part@20} \\
\midrule
\texttt{Naive}
& 67.11 & 90.06 & 96.45 & 98.34
& 32.92 & 89.23 & 99.85 & 100.00
& 8.00 & 22.50 & 29.8 & 50.7 \\
\midrule
\texttt{1 Question}
& 66.02 & 90.12 & 96.07 & 98.58
& 32.46 & 91.85 & 100.00 & 100.00
& 15.00 & 30.00 & 37.0 & 61.2 \\
\texttt{5 Questions}
& 77.05 & 94.90 & 97.81 & \textbf{99.24}
& 35.85 & 92.46 & 99.69 & 100.00
& 12.00 & 33.00 & 37.9 & 61.6 \\
\texttt{10 Questions}
& 77.58 & \textbf{95.03} & 97.39 & 98.41
& 35.85 & \textbf{92.46} & 99.54 & 100.00
& 16.00 & 35.00 & 38.2 & 62.2 \\
\texttt{15 Questions}
& 77.22 & 94.05 & 96.49 & 97.47
& \textbf{38.31} & 90.92 & 99.69 & 100.00
& 14.00 & 34.00 & 32.1 & 60.3 \\
\texttt{20 Questions}
& 76.24 & 93.07 & 95.73 & 96.45
& 35.85 & 92.31 & 99.69 & 100.00
& 14.00 & \textbf{35.00} & 34.9 & \textbf{63.8} \\
\texttt{30 Questions}
& 76.05 & 92.18 & 95.05 & 95.73
& 34.00 & 90.15 & 99.54 & 99.69
& 11.00 & 30.00 & 31.5 & 58.5 \\
\midrule
\texttt{LLM Decides}
& \textbf{77.65} & 93.97 & \textbf{96.30} & 97.13
& 38.00 & 92.15 & 99.69 & 100.00
& \textbf{18.00} & \textbf{35.00} & \textbf{41.9} & 62.7 \\
\bottomrule
\end{tabular}
\end{table*}

We observe a similar pattern in \emph{Natural Questions} (Table~\ref{tab:questions-comparison}). The naive approach 
(and letting the LLM decide automatically) both hover around 61–64\% Top-1 Context Accuracy. 
Generating 15 or 20 questions per chunk can push Top-1 Context Accuracy slightly higher, surpassing 64\%. 
In general, \emph{Title Accuracy} improves more noticeably, approaching or exceeding 79\% at Top-1 when 15+ questions are used.

\begin{itemize}
    \item \textbf{Moderate Gains.} Unlike SQuAD, the gains from adding more questions in NQ are more modest 
    (e.g., from 61\% to 65\% in Top-1 Context Accuracy).
    \item \textbf{Title Accuracy.} By contrast, Title Accuracy climbs above 79\% at Top-1 (with 15–20 questions), 
    indicating that question generation consistently helps the system find the right \emph{article}, even if 
    the precise paragraph-level retrieval remains challenging.
\end{itemize}

\paragraph{MultiHop-RAG}
Because MultiHop-RAG tasks require retrieving \emph{all relevant documents}, we track both \emph{Full Match Accuracy} 
and \emph{Partial Match Statistics}. As seen in Table~\ref{tab:questions-comparison}:

\begin{itemize}
    \item \textbf{Full Match Accuracy.} Baseline naive retrieval (LLM Naive) achieves only about 10\% at $k=5$ 
    and 37\% at $k=20$. Using question generation with 5 or 10 questions can improve the $k=5$ Full Match 
    from 10–12\% to 15–16\%, and from 19\% to 24–25\% at $k=10$. Even at higher $k$ values (15 or 20), 
    best-case Full Match remains in the 30–35\% range, underscoring the challenge of truly multi-hop retrieval. 
    Interestingly, letting the LLM decide (without specifying a question count) yields 18\% at $k=5$ and 35\% at $k=20$.
    \item \textbf{Partial Match.} We also evaluate the \emph{average percentage of required evidence} retrieved. 
    Generating around 5–20 questions consistently pushes partial match rates above 50–60\% at $k=15$ or $k=20$, 
    compared to 35–45\% with naive retrieval. This indicates that even when the system does not achieve a complete 
    full match, it still locates \emph{some} of the essential documents for partial reasoning.
\end{itemize}

These findings confirm that, while \emph{multi-hop} queries remain significantly harder, carefully chosen question sets 
(i.e., 10–20 questions) yield noticeable improvements over a naive approach.

\begin{itemize}
    \item Generating more questions typically improves retrieval performance, but returns diminish beyond 
    about \emph{10–15 questions per chunk}.
    \item Even a moderate number of questions (5–10) can outperform naive retrieval by a wide margin in both 
    single-hop (NQ, SQuAD) and multi-hop (RAG) settings.
    \item In \emph{multi-hop} scenarios, \emph{partial match} is also improved by question generation, 
    indicating that the system at least retrieves some relevant documents more reliably.
\end{itemize}

Overall, most datasets show a \emph{sweet spot} around 10–15 questions, balancing coverage with potential redundancy. 
Although letting the LLM fully decide the number of questions can yield strong results in certain cases (e.g., SQuAD), 
the performance varies by dataset. Consequently, the optimal question count appears to depend on domain complexity 
and the specifics of the QA task.

\subsection{Comparison with HyDE}
\label{subsec:compare-hyde}

A popular technique for query enrichment is \textbf{HyDE}, which generates a hypothetical document at \emph{query time} before embedding it and retrieving relevant chunks. Although HyDE can improve coverage, it requires an LLM call for each incoming query, introducing significant latency. In contrast, \textbf{QuOTE} moves question generation to \emph{index time}, incurring a one-time cost but speeding up the overall querying process. We compare \textbf{Naive RAG} (no query transformations), \textbf{HyDE}, and \textbf{QuOTE} on all three benchmarks. For \textbf{SQuAD} and \textbf{NQ} we showcase
\emph{Top-1 Context Accuracy}
and for \textbf{MultiHop-RAG (multi-hop)}, because queries need multiple pieces of evidence, we focus on \emph{Full Match} at \(k=20\). 

Table~\ref{tab:hyde-comparison} shows that while HyDE sometimes boosts accuracy compared to a Naive approach, its {\bf per-query LLM calls lead to a drastic rise in average retrieval time} (\(\mathbf{+1}\text{--}2 \text{ seconds}\) per query). By contrast, QuOTE often equals or surpasses Naive’s retrieval accuracy with only a modest query-time overhead, as most of its work is done in the indexing phase.

Overall, these findings illustrate that while \textbf{HyDE} can be valuable in certain multi-hop or complex queries, it incurs a substantial latency cost. 
In fact, \textbf{HyDE} can be viewed as primarily a {\bf test-time compute} innovation.
\textbf{QuOTE} offers significant advantages: higher retrieval accuracy than Naive in most cases, a one-time, amortized cost for question generation, and dramatic query latency improvements over HyDE.

\begin{table*}[t]
\centering
\caption{Comparison of \textbf{Naive}, \textbf{HyDE}, and \textbf{QuOTE} across three QA tasks. 
The \textbf{fastest} (lowest time) and \textbf{most accurate} (highest accuracy) entries in each column are \textbf{bolded}. 
In \textbf{SQuAD}, \texttt{Naive} is fastest while \texttt{QuOTE} achieves the highest accuracy; 
for \textbf{NQ}, \texttt{Naive} runs fastest while \texttt{HyDE} slightly outperforms the others in accuracy; 
and in \textbf{MultiHop-RAG}, \texttt{Naive} remains fastest, whereas \texttt{QuOTE} attains the highest full-match rates.}
\label{tab:hyde-comparison}
\renewcommand{\arraystretch}{1.15}
\setlength{\tabcolsep}{3pt}
\begin{tabular}{l|cccc|cccc|cccc}
\toprule
& \multicolumn{4}{c|}{\textbf{SQuAD}} 
& \multicolumn{4}{c|}{\textbf{NQ}} 
& \multicolumn{4}{c}{\textbf{MultiHop-RAG}} \\
\cmidrule(lr){2-5}\cmidrule(lr){6-9}\cmidrule(lr){10-13}
\textbf{Approach} 
& \textbf{Time(s)} & \textbf{ms/q} & \textbf{C@1} & \textbf{C@5}
& \textbf{Time(s)} & \textbf{ms/q} & \textbf{C@1} & \textbf{C@5}
& \textbf{Time(s)} & \textbf{ms/q} & \textbf{Full@5} & \textbf{Full@20} \\
\midrule
\textbf{Naive} & 
\textbf{99.97} & \textbf{108.31} & 79.31\% & 95.88\% & 
\textbf{198.96} & \textbf{187.34} & 32.92\% & 89.23\% & 
\textbf{3.10} & \textbf{15.14} & 7.00\% & 23.00\% \\
\textbf{HyDE} &
1176.20 & 1274.32 & 76.60\% & 92.63\% & 
2707.56 & 2549.49 & 33.23\% & 90.46\% &
1155.80 & 4157.83 & 6.00\% & 23.50\% \\
\textbf{QuOTE} &
130.56 & 141.46 & \textbf{90.03}\% & \textbf{98.48}\% & 
388.38 & 365.71 & \textbf{38.00}\% & \textbf{92.15}\% &
926.41 & 100.75 & \textbf{8.00}\% & \textbf{29.00}\% \\
\bottomrule
\end{tabular}
\end{table*}

\subsection{Effect of Deduplication}
\label{subsec:dedup-effect}

Deduplication is essential in \textbf{QuOTE} because each chunk can be indexed multiple times---once per generated question---leading to redundant matches at query time. Again, We compare \textbf{Naive RAG},
\textbf{HyDE} and
\textbf{QuOTE}.
When $k=1$, deduplication is unnecessary, as only one chunk is retrieved. However, when $k \in \{5,10,20\}$, \texttt{QuOTE} systems 
fetch more than $k$ results from the vector index (e.g., $k \times 5$) 
and then deduplicate by original chunk text. This extra step 
introduces a small overhead, but we find that QuOTE remains much faster than HyDE (which invokes an LLM at \emph{each} query) 
and substantially outperforms Naive in Top-1 Context Accuracy.

Table~\ref{tab:dedup-quote} shows a head-to-head comparison of the three approaches on a SQuAD subset. 
Each approach processes 923 queries 
for all benchmarks, we see that
\textbf{Naive} is the fastest and \textbf{QuOTE} the most accurate.
The added overhead incurred by \textbf{QuOTE} over \textbf{Naive}
small relative to the cost of \emph{per-query} generation in HyDE. 
Hence, \textbf{QuOTE} obtains both \emph{superior accuracy} and \emph{faster query times} than HyDE, 
while incurring a one-time cost for indexing. In settings where repeated queries are common, 
paying a higher index-time cost can significantly improve responsiveness and end-user experience.

\begin{table}[t]
\centering
\caption{Comparison of retrieval approaches. Index=time to build database (seconds), Query=time to process all queries (seconds), ms/q=milliseconds per query, C@1=Context accuracy, T@1=Title accuracy. Bold indicates best per column.}
\label{tab:dedup-quote}
\renewcommand{\arraystretch}{1.05}
\begin{tabular}{lccccc}
\toprule
\textbf{App} & \textbf{Index} & \textbf{Query} & \textbf{ms/q} & \textbf{C@1} & \textbf{T@1} \\
\midrule
Naive & 10.36 & \textbf{99.97} & \textbf{108.31} & 79.31\% & 96.64\% \\
HyDE & \textbf{10.14} & 1176.20 & 1274.32 & 76.60\% & 94.58\% \\
QuOTE & 170.32 & 130.56 & 141.46 & \textbf{90.03\%} & \textbf{98.37\%} \\
\bottomrule
\end{tabular}
\end{table}


\subsection{Can we use a Cheaper LLM for Question Generation?}
\label{subsec:cheaper-models}
An important practical consideration in RAG-based pipelines is whether \emph{cheaper, smaller models} can generate 
effective questions for indexing, or if premium, large-scale LLMs (e.g., GPT-4) are necessary. To investigate, 
we experimented with a variety of local language models (e.g., \texttt{gemma2-9b}, \texttt{llama3-8b}, and \texttt{qwen2.5-7b}), 
as well as \texttt{gpt-4o-mini}, \texttt{gpt-4o}, and a baseline \texttt{Naive} approach that relies solely on 
the chunk text without question generation. All runs were conducted on a \emph{SQuAD-based subset}. Table~\ref{tab:cheaper-models} 
summarizes the results in terms of 
\textbf{Top-$k$ Context Accuracy} and \textbf{Top-$k$ Title Accuracy}.

\begin{table}[t]
\centering
\caption{Comparison of different models on a SQuAD subset. We report Context Accuracy (C@k) and Title Accuracy (T@k) at k=1 and k=5. Best value(s) in each column are \textbf{bolded}.}
\label{tab:cheaper-models}
\renewcommand{\arraystretch}{1.1}
\begin{tabular}{l|cc|cc}
\toprule
\textbf{Model} & \textbf{C@1} & \textbf{C@5} & \textbf{T@1} & \textbf{T@5} \\
\midrule
\texttt{Naive}       
& 66.60 & 92.80 & 98.32 & 99.44 \\
\midrule
\texttt{QuOTE gemma2-9b}   
& 74.49 & 96.93 & \textbf{99.09} & 99.72 \\
\texttt{QuOTE gpt-4o-mini} 
& \textbf{76.73} & \textbf{97.20} & 98.81 & 99.30 \\
\texttt{QuOTE gpt-4o}      
& 76.38 & 96.72 & \textbf{99.09} & 99.58 \\
\texttt{QuOTE llama3-8b}   
& 73.58 & 95.95 & \textbf{99.09} & \textbf{99.79} \\
\texttt{QuOTE llama3.1-8b} 
& 71.42 & 96.09 & \textbf{99.09} & 99.72 \\
\texttt{QuOTE llama3.2-3b} 
& 70.86 & 94.90 & 98.81 & 99.65 \\
\texttt{QuOTE phi4} 
& 74.07 & 95.95 & 98.81 & 99.30 \\
\texttt{QuOTE qwen2.5-7b}  
& 72.05 & 96.23 & 99.02 & 99.72 \\
\bottomrule
\end{tabular}
\end{table}

We observe that even
smaller models such as 
    \texttt{llama3.2-3b} achieve over 70\% Top-1 Context Accuracy---only a few percentage points behind the more capable 
    \texttt{gpt-4o-mini} or \texttt{gpt-4o} models.
For nearly all models, \emph{Top-1 Title Accuracy} remains around 
    or above 98\%, indicating that the question generation step—regardless of the model size—helps QuOTE 
hone in on the correct article.
Once we allow for more retrieved chunks 
    (Top-10 or Top-20), nearly all approaches exceed 98--99\% Context Accuracy. This suggests that 
    question augmentation significantly reduces misalignment with relevant passages.
    The main trade-off is that \texttt{gpt-4o} and \texttt{gpt-4o-mini} 
    exhibit slightly higher Top-1 Context Accuracy (up to $\sim76\%$) compared to cheaper models 
    (70--74\%); however, local LLMs still offer near-parity in mid- to high-$k$ retrieval settings 
    with notably lower inference cost.

\subsection{Effect of the Number of Contexts on Retrieval Accuracy}
\label{subsec:context-vs-accuracy}

Analysis of the impact of the number of contexts on retrieval performance reveals distinct patterns between SQuAD and NQ, reflecting their fundamentally different dataset characteristics.

\begin{figure}[ht]
\centering
\includegraphics[width=0.8\columnwidth]{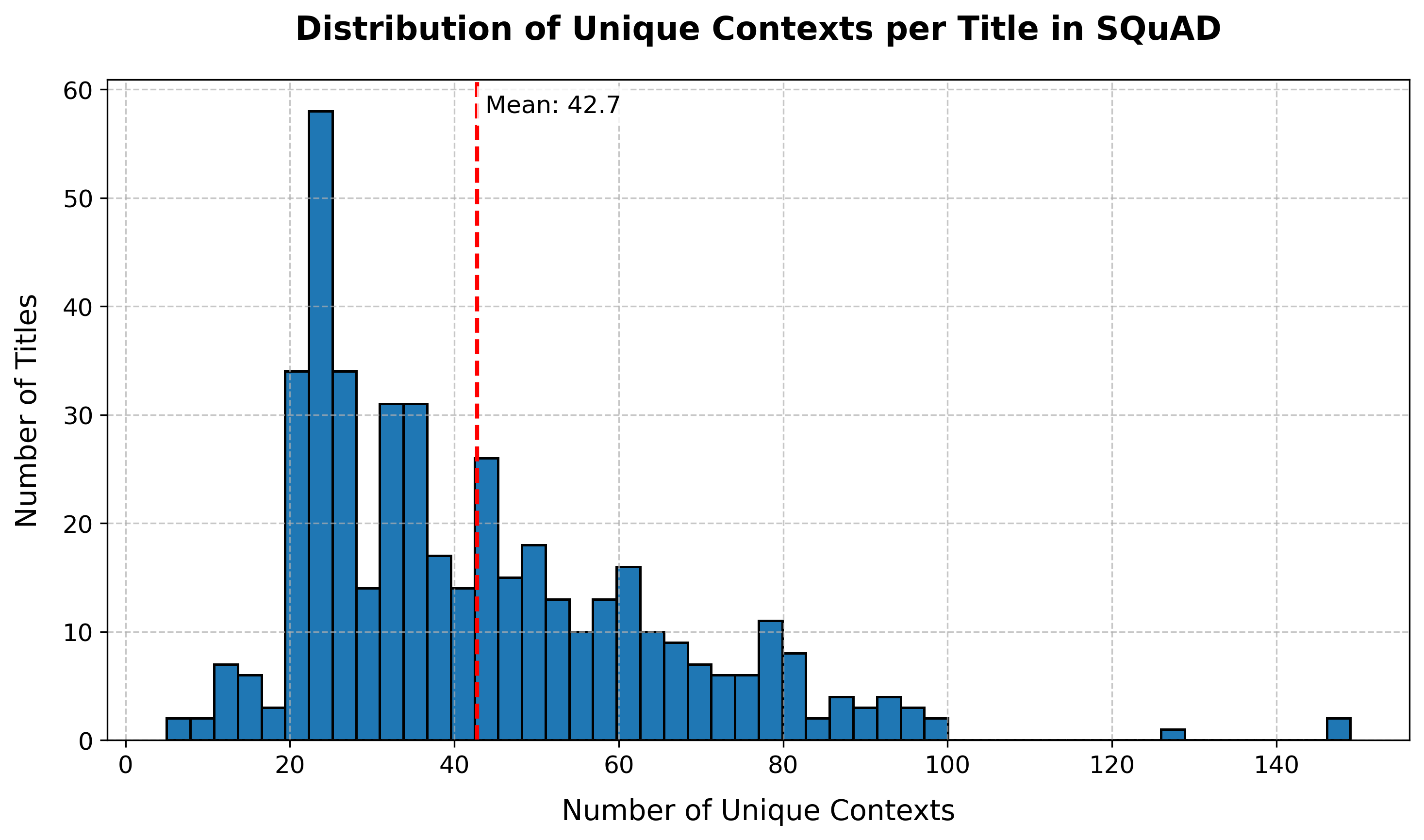}
\caption{Distribution of contexts per title in SQuAD (N=442 titles). The mean of 42.74 contexts per title and maximum of 149 contexts demonstrate the dataset's high context density.}
\label{fig:squad_distribution}
\end{figure}

SQuAD exhibits a rich context structure, with 442 titles having a mean of 42.74 contexts per title (median=36). This substantial density, ranging from 5 to 149 contexts per title, creates significant potential for confusion with naive retrieval approaches, particularly when similar passages exist within the same document.

\begin{figure}[ht]
\centering
\includegraphics[width=0.8\columnwidth]{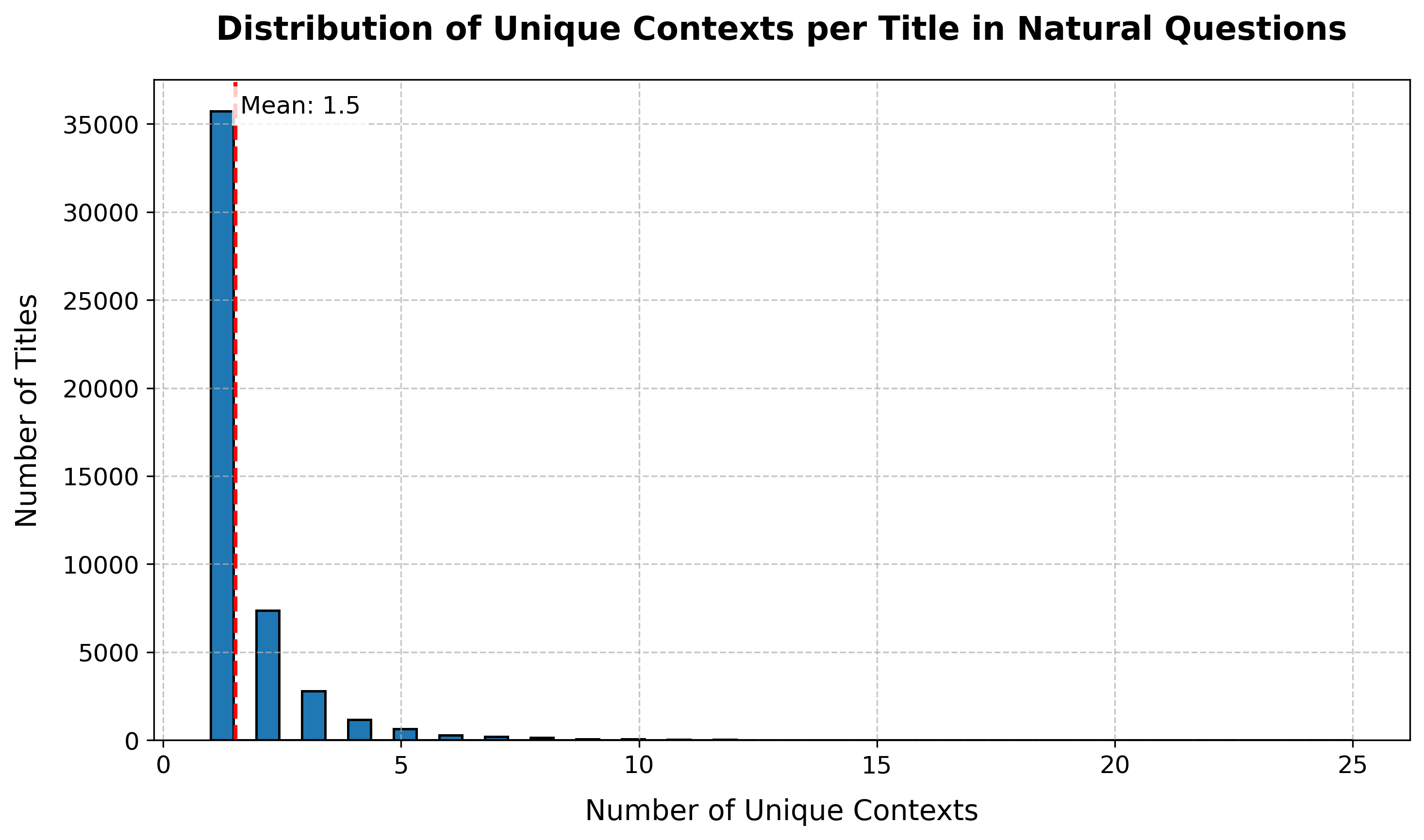}
\caption{Distribution of contexts per title in Natural Questions (N=48,525 titles). The highly concentrated distribution around a median of 1 context per title indicates predominantly singular contexts.}
\label{fig:nq_distribution}
\end{figure}

In stark contrast, NQ presents a much sparser context landscape. Across its 48,525 titles, NQ maintains a mean of just 1.52 contexts per title, with a median of 1, indicating that most titles have unique contexts.

\begin{figure}[ht]
\centering
\includegraphics[width=1\columnwidth]{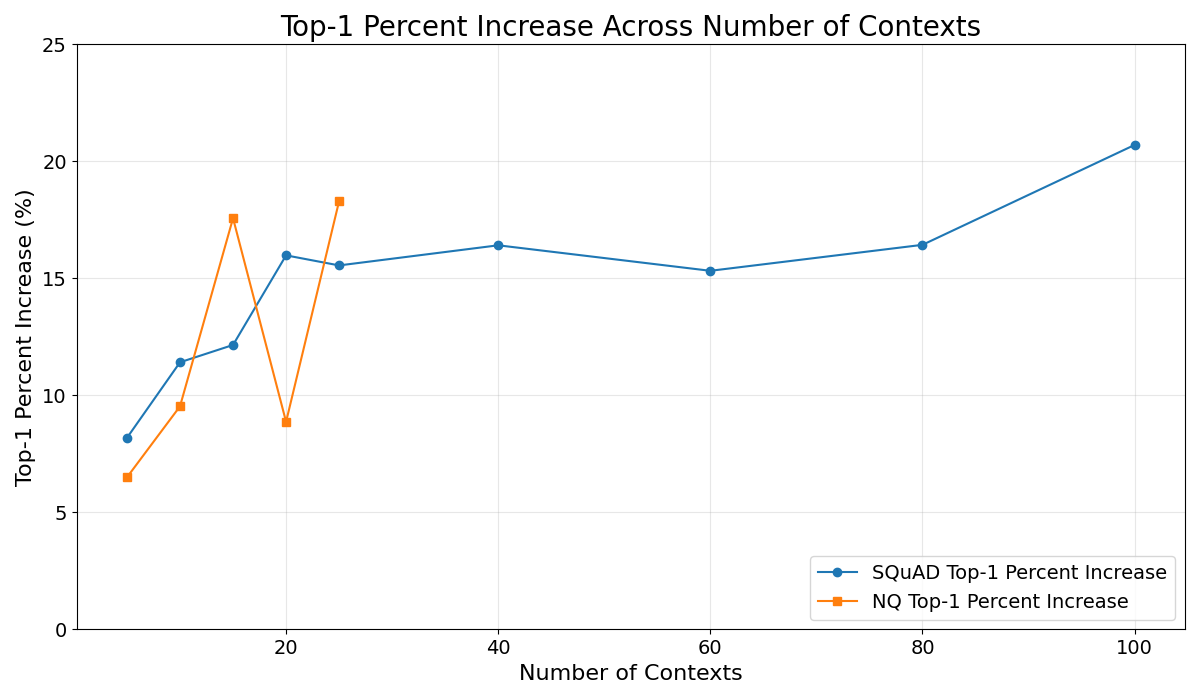}
\caption{Percentage increase in Top-1 retrieval accuracy with QuOTE compared to naive retrieval across the number of contexts. SQuAD shows steady improvement that grows with the number of contexts, reaching 20.7\% improvement at size 100, while NQ shows consistent but variable gains up to 18.3\%.}
\label{fig:improvement}
\end{figure}

These structural differences manifest clearly in QuOTE's relative performance gains (Figure \ref{fig:improvement}). For SQuAD, we observe a steady increase in QuOTE's advantage as the number of contexts grows. Starting from around 8\% improvement in a small number of contexts, it increases consistently to about 16\% at moderate number of contexts, and continues to improve to exceed the improvement 20\% for a larger number of contexts. This steady improvement trend aligns with the increasing challenge of disambiguating similar contexts in longer documents.

NQ shows a more constrained but generally positive pattern, reflecting its simpler context structure. The improvements start at about 6\% for a small number of contexts and reach peaks of approximately 18\% for a moderate number of contexts. Although the magnitude of the improvements varies, QuOTE consistently enhances the retrieval accuracy in most contexts, although its impact is more variable than in SQuAD.

These insights demonstrate how dataset characteristics fundamentally influence QuOTE's effectiveness. For collections with many contexts per document like SQuAD, QuOTE provides increasingly valuable disambiguation as document length grows. For collections like NQ where most documents contain just a single relevant chunk, QuOTE still provides consistent benefits, though the magnitude varies with the number of contexts.
\section{Discussion}
\label{sec:discussion}
This work has demonstrated how the use of questions to augment representations of documents can yield significant improvement in information retrieval for RAG applications. The need for deduplication introduced by our approach does not incur a significant overhead and can instead improve retrieval quality across a range of benchmarks. 

There are several possible directions of future work. One promising direction is the development of a \emph{self-improving} indexing strategy, possibly with an LLM fine-tuning approach, that adapts over time. 
Specifically, we could monitor user queries and their corresponding feedback (e.g., whether the user found the retrieved context helpful) and selectively ingest \emph{new or corrected} query--context pairs into the index. 

A second direction of future research involves developing \emph{prompt optimization} frameworks (e.g., through automated prompt search or via reinforcement learning) to improve question-generation quality. 
By systematically tuning prompts, we may generate more precise and context-rich questions for each chunk. 

Finally, we can imagine embedding some documents as-is and others with our augmented questions, developing a hybrid approach to RAG. To support the design of such systems, we intend to explore the development of scaling laws w.r.t. all the parameters studied here.
%


\bibliographystyle{ACM-Reference-Format}
\bibliography{reference}


\begin{thebibliography}{41}


\ifx \showCODEN    \undefined \def \showCODEN     #1{\unskip}     \fi
\ifx \showDOI      \undefined \def \showDOI       #1{#1}\fi
\ifx \showISBNx    \undefined \def \showISBNx     #1{\unskip}     \fi
\ifx \showISBNxiii \undefined \def \showISBNxiii  #1{\unskip}     \fi
\ifx \showISSN     \undefined \def \showISSN      #1{\unskip}     \fi
\ifx \showLCCN     \undefined \def \showLCCN      #1{\unskip}     \fi
\ifx \shownote     \undefined \def \shownote      #1{#1}          \fi
\ifx \showarticletitle \undefined \def \showarticletitle #1{#1}   \fi
\ifx \showURL      \undefined \def \showURL       {\relax}        \fi
\providecommand\bibfield[2]{#2}
\providecommand\bibinfo[2]{#2}
\providecommand\natexlab[1]{#1}
\providecommand\showeprint[2][]{arXiv:#2}

\bibitem[ant({[n.\,d.]})]%
        {anthropic_contextual_retrieval}
 \bibinfo{year}{[n.\,d.]}\natexlab{}.
\newblock \bibinfo{title}{Anthropic Contextual Retrieval}.
\newblock \bibinfo{howpublished}{\url{https://www.anthropic.com/news/contextual-retrieval}}.
\newblock


\bibitem[Anantha et~al\mbox{.}(2023)]%
        {rag-variant3}
\bibfield{author}{\bibinfo{person}{Raviteja Anantha}, \bibinfo{person}{Tharun Bethi}, \bibinfo{person}{Danil Vodianik}, {and} \bibinfo{person}{Srinivas Chappidi}.} \bibinfo{year}{2023}\natexlab{}.
\newblock \showarticletitle{Context tuning for retrieval augmented generation}.
\newblock \bibinfo{journal}{\emph{arXiv preprint arXiv:2312.05708}} (\bibinfo{year}{2023}).
\newblock


\bibitem[Chen et~al\mbox{.}(2021)]%
        {sparse-vs-dense1}
\bibfield{author}{\bibinfo{person}{Xilun Chen}, \bibinfo{person}{Kushal Lakhotia}, \bibinfo{person}{Barlas O{\u{g}}uz}, \bibinfo{person}{Anchit Gupta}, \bibinfo{person}{Patrick Lewis}, \bibinfo{person}{Stan Peshterliev}, \bibinfo{person}{Yashar Mehdad}, \bibinfo{person}{Sonal Gupta}, {and} \bibinfo{person}{Wen-tau Yih}.} \bibinfo{year}{2021}\natexlab{}.
\newblock \showarticletitle{Salient phrase aware dense retrieval: can a dense retriever imitate a sparse one?}
\newblock \bibinfo{journal}{\emph{arXiv preprint arXiv:2110.06918}} (\bibinfo{year}{2021}).
\newblock


\bibitem[Cheng et~al\mbox{.}(2024)]%
        {rag-variant2}
\bibfield{author}{\bibinfo{person}{Xin Cheng}, \bibinfo{person}{Xun Wang}, \bibinfo{person}{Xingxing Zhang}, \bibinfo{person}{Tao Ge}, \bibinfo{person}{Si-Qing Chen}, \bibinfo{person}{Furu Wei}, \bibinfo{person}{Huishuai Zhang}, {and} \bibinfo{person}{Dongyan Zhao}.} \bibinfo{year}{2024}\natexlab{}.
\newblock \showarticletitle{xRAG: Extreme Context Compression for Retrieval-augmented Generation with One Token}.
\newblock \bibinfo{journal}{\emph{arXiv preprint arXiv:2405.13792}} (\bibinfo{year}{2024}).
\newblock


\bibitem[Cuconasu et~al\mbox{.}(2024)]%
        {cuconasu}
\bibfield{author}{\bibinfo{person}{Florin Cuconasu}, \bibinfo{person}{Giovanni Trappolini}, \bibinfo{person}{Federico Siciliano}, \bibinfo{person}{Simone Filice}, \bibinfo{person}{Cesare Campagnano}, \bibinfo{person}{Yoelle Maarek}, \bibinfo{person}{Nicola Tonellotto}, {and} \bibinfo{person}{Fabrizio Silvestri}.} \bibinfo{year}{2024}\natexlab{}.
\newblock \showarticletitle{The power of noise: Redefining retrieval for rag systems}. In \bibinfo{booktitle}{\emph{Proceedings of the 47th International ACM SIGIR Conference on Research and Development in Information Retrieval}}. \bibinfo{pages}{719--729}.
\newblock


\bibitem[Fan et~al\mbox{.}(2019)]%
        {eli5}
\bibfield{author}{\bibinfo{person}{Angela Fan}, \bibinfo{person}{Yacine Jernite}, \bibinfo{person}{Ethan Perez}, \bibinfo{person}{David Grangier}, \bibinfo{person}{Jason Weston}, {and} \bibinfo{person}{Michael Auli}.} \bibinfo{year}{2019}\natexlab{}.
\newblock \showarticletitle{ELI5: Long form question answering}.
\newblock \bibinfo{journal}{\emph{arXiv preprint arXiv:1907.09190}} (\bibinfo{year}{2019}).
\newblock


\bibitem[Gao et~al\mbox{.}(2022)]%
        {hyde2022}
\bibfield{author}{\bibinfo{person}{Luyu Gao}, \bibinfo{person}{Xueguang Ma}, \bibinfo{person}{Jimmy Lin}, {and} \bibinfo{person}{Jamie Callan}.} \bibinfo{year}{2022}\natexlab{}.
\newblock \showarticletitle{Precise Zero-Shot Dense Retrieval without Relevance Labels}.
\newblock \bibinfo{journal}{\emph{arXiv preprint arXiv:2212.10496}} (\bibinfo{year}{2022}).
\newblock


\bibitem[Gim et~al\mbox{.}(2024)]%
        {prompt-caching-RAG}
\bibfield{author}{\bibinfo{person}{In Gim}, \bibinfo{person}{Guojun Chen}, \bibinfo{person}{Seung-seob Lee}, \bibinfo{person}{Nikhil Sarda}, \bibinfo{person}{Anurag Khandelwal}, {and} \bibinfo{person}{Lin Zhong}.} \bibinfo{year}{2024}\natexlab{}.
\newblock \showarticletitle{Prompt cache: Modular attention reuse for low-latency inference}.
\newblock \bibinfo{journal}{\emph{Proceedings of Machine Learning and Systems}}  \bibinfo{volume}{6} (\bibinfo{year}{2024}), \bibinfo{pages}{325--338}.
\newblock


\bibitem[Heilman and Smith(2009)]%
        {heilman_smith_2009}
\bibfield{author}{\bibinfo{person}{Michael Heilman} {and} \bibinfo{person}{Noah~A. Smith}.} \bibinfo{year}{2009}\natexlab{}.
\newblock \showarticletitle{Ranking Automatically Generated Questions as a Shared Task}. In \bibinfo{booktitle}{\emph{Proceedings of the AIED Workshop on Question Generation}}. \bibinfo{address}{Brighton, UK}.
\newblock


\bibitem[Heilman and Smith(2010)]%
        {heilman_smith_2010}
\bibfield{author}{\bibinfo{person}{Michael Heilman} {and} \bibinfo{person}{Noah~A. Smith}.} \bibinfo{year}{2010}\natexlab{}.
\newblock \showarticletitle{Good Question! Statistical Ranking for Question Generation}. In \bibinfo{booktitle}{\emph{Proceedings of the North American Chapter of the Association for Computational Linguistics (NAACL)}}. \bibinfo{address}{Los Angeles, CA}.
\newblock


\bibitem[Indyk and Motwani(1998)]%
        {ann-ir-paper}
\bibfield{author}{\bibinfo{person}{Piotr Indyk} {and} \bibinfo{person}{Rajeev Motwani}.} \bibinfo{year}{1998}\natexlab{}.
\newblock \showarticletitle{Approximate nearest neighbors: towards removing the curse of dimensionality}. In \bibinfo{booktitle}{\emph{Proceedings of the thirtieth annual ACM symposium on Theory of computing}}. \bibinfo{pages}{604--613}.
\newblock


\bibitem[Jacob et~al\mbox{.}(2024)]%
        {Jacob-Drozdov-drowning-in-documents}
\bibfield{author}{\bibinfo{person}{Mathew Jacob}, \bibinfo{person}{Erik Lindgren}, \bibinfo{person}{Matei Zaharia}, \bibinfo{person}{Michael Carbin}, \bibinfo{person}{Omar Khattab}, {and} \bibinfo{person}{Andrew Drozdov}.} \bibinfo{year}{2024}\natexlab{}.
\newblock \showarticletitle{Drowning in Documents: Consequences of Scaling Reranker Inference}.
\newblock \bibinfo{journal}{\emph{arXiv preprint arXiv:2411.11767}} (\bibinfo{year}{2024}).
\newblock


\bibitem[Jiang et~al\mbox{.}(2024)]%
        {rag-variant1}
\bibfield{author}{\bibinfo{person}{Ziyan Jiang}, \bibinfo{person}{Xueguang Ma}, {and} \bibinfo{person}{Wenhu Chen}.} \bibinfo{year}{2024}\natexlab{}.
\newblock \showarticletitle{Longrag: Enhancing retrieval-augmented generation with long-context llms}.
\newblock \bibinfo{journal}{\emph{arXiv preprint arXiv:2406.15319}} (\bibinfo{year}{2024}).
\newblock


\bibitem[Joshi et~al\mbox{.}(2017)]%
        {triviaqa}
\bibfield{author}{\bibinfo{person}{Mandar Joshi}, \bibinfo{person}{Eunsol Choi}, \bibinfo{person}{Daniel~S Weld}, {and} \bibinfo{person}{Luke Zettlemoyer}.} \bibinfo{year}{2017}\natexlab{}.
\newblock \showarticletitle{Triviaqa: A large scale distantly supervised challenge dataset for reading comprehension}.
\newblock \bibinfo{journal}{\emph{arXiv preprint arXiv:1705.03551}} (\bibinfo{year}{2017}).
\newblock


\bibitem[Khanda(2024)]%
        {ragc}
\bibfield{author}{\bibinfo{person}{Rajat Khanda}.} \bibinfo{year}{2024}\natexlab{}.
\newblock \showarticletitle{Agentic AI-Driven Technical Troubleshooting for Enterprise Systems: A Novel Weighted Retrieval-Augmented Generation Paradigm}.
\newblock \bibinfo{journal}{\emph{arXiv preprint arXiv:2412.12006}} (\bibinfo{year}{2024}).
\newblock


\bibitem[Khattab and Zaharia(2020)]%
        {colbert-3}
\bibfield{author}{\bibinfo{person}{Omar Khattab} {and} \bibinfo{person}{Matei Zaharia}.} \bibinfo{year}{2020}\natexlab{}.
\newblock \showarticletitle{Colbert: Efficient and effective passage search via contextualized late interaction over bert}. In \bibinfo{booktitle}{\emph{Proceedings of the 43rd International ACM SIGIR conference on research and development in Information Retrieval}}. \bibinfo{pages}{39--48}.
\newblock


\bibitem[Kwiatkowski et~al\mbox{.}(2019)]%
        {nq}
\bibfield{author}{\bibinfo{person}{Tom Kwiatkowski}, \bibinfo{person}{Jennimaria Palomaki}, \bibinfo{person}{Olivia Redfield}, \bibinfo{person}{Michael Collins}, \bibinfo{person}{Ankur Parikh}, \bibinfo{person}{Chris Alberti}, \bibinfo{person}{Danielle Epstein}, \bibinfo{person}{Illia Polosukhin}, \bibinfo{person}{Jacob Devlin}, \bibinfo{person}{Kenton Lee}, {et~al\mbox{.}}} \bibinfo{year}{2019}\natexlab{}.
\newblock \showarticletitle{Natural questions: a benchmark for question answering research}.
\newblock \bibinfo{journal}{\emph{Transactions of the Association for Computational Linguistics}}  \bibinfo{volume}{7} (\bibinfo{year}{2019}), \bibinfo{pages}{453--466}.
\newblock


\bibitem[Leto et~al\mbox{.}(2024)]%
        {leto2024toward}
\bibfield{author}{\bibinfo{person}{Alexandria Leto}, \bibinfo{person}{Cecilia Aguerrebere}, \bibinfo{person}{Ishwar Bhati}, \bibinfo{person}{Ted Willke}, \bibinfo{person}{Mariano Tepper}, {and} \bibinfo{person}{Vy~Ai Vo}.} \bibinfo{year}{2024}\natexlab{}.
\newblock \showarticletitle{Toward Optimal Search and Retrieval for RAG}.
\newblock \bibinfo{journal}{\emph{arXiv preprint arXiv:2411.07396}} (\bibinfo{year}{2024}).
\newblock


\bibitem[Li et~al\mbox{.}(2023)]%
        {multi-hop-7}
\bibfield{author}{\bibinfo{person}{Jiawei Li}, \bibinfo{person}{Mucheng Ren}, \bibinfo{person}{Yang Gao}, {and} \bibinfo{person}{Yizhe Yang}.} \bibinfo{year}{2023}\natexlab{}.
\newblock \showarticletitle{Ask to Understand: Question Generation for Multi-hop Question Answering}. In \bibinfo{booktitle}{\emph{China National Conference on Chinese Computational Linguistics}}. Springer, \bibinfo{pages}{19--36}.
\newblock


\bibitem[Liu et~al\mbox{.}(2021)]%
        {raga}
\bibfield{author}{\bibinfo{person}{Ye Liu}, \bibinfo{person}{Kazuma Hashimoto}, \bibinfo{person}{Yingbo Zhou}, \bibinfo{person}{Semih Yavuz}, \bibinfo{person}{Caiming Xiong}, {and} \bibinfo{person}{Philip~S Yu}.} \bibinfo{year}{2021}\natexlab{}.
\newblock \showarticletitle{Dense hierarchical retrieval for open-domain question answering}.
\newblock \bibinfo{journal}{\emph{arXiv preprint arXiv:2110.15439}} (\bibinfo{year}{2021}).
\newblock


\bibitem[Malkov et~al\mbox{.}(2014)]%
        {anncomp2}
\bibfield{author}{\bibinfo{person}{Yury Malkov}, \bibinfo{person}{Alexander Ponomarenko}, \bibinfo{person}{Andrey Logvinov}, {and} \bibinfo{person}{Vladimir Krylov}.} \bibinfo{year}{2014}\natexlab{}.
\newblock \showarticletitle{Approximate nearest neighbor algorithm based on navigable small world graphs}.
\newblock \bibinfo{journal}{\emph{Information Systems}}  \bibinfo{volume}{45} (\bibinfo{year}{2014}), \bibinfo{pages}{61--68}.
\newblock


\bibitem[Mavi et~al\mbox{.}(2022)]%
        {mavi2022multihop}
\bibfield{author}{\bibinfo{person}{Vaibhav Mavi}, \bibinfo{person}{Anubhav Jangra}, {and} \bibinfo{person}{Adam Jatowt}.} \bibinfo{year}{2022}\natexlab{}.
\newblock \showarticletitle{Multi-hop Question Answering}.
\newblock \bibinfo{journal}{\emph{arXiv preprint arXiv:2204.09140}} (\bibinfo{year}{2022}).
\newblock
\urldef\tempurl%
\url{https://doi.org/10.48550/arXiv.2204.09140}
\showURL{%
\tempurl}
\newblock
\shownote{Published at Foundations and Trends in Information Retrieval}.


\bibitem[Rajpurkar(2016)]%
        {squad1}
\bibfield{author}{\bibinfo{person}{P Rajpurkar}.} \bibinfo{year}{2016}\natexlab{}.
\newblock \showarticletitle{Squad: 100,000+ questions for machine comprehension of text}.
\newblock \bibinfo{journal}{\emph{arXiv preprint arXiv:1606.05250}} (\bibinfo{year}{2016}).
\newblock


\bibitem[Rajpurkar et~al\mbox{.}(2018)]%
        {squad2}
\bibfield{author}{\bibinfo{person}{Pranav Rajpurkar}, \bibinfo{person}{Robin Jia}, {and} \bibinfo{person}{Percy Liang}.} \bibinfo{year}{2018}\natexlab{}.
\newblock \showarticletitle{Know what you don't know: Unanswerable questions for SQuAD}.
\newblock \bibinfo{journal}{\emph{arXiv preprint arXiv:1806.03822}} (\bibinfo{year}{2018}).
\newblock


\bibitem[Rangan and Yin(2024)]%
        {ragimp2}
\bibfield{author}{\bibinfo{person}{Keshav Rangan} {and} \bibinfo{person}{Yiqiao Yin}.} \bibinfo{year}{2024}\natexlab{}.
\newblock \showarticletitle{A fine-tuning enhanced RAG system with quantized influence measure as AI judge}.
\newblock \bibinfo{journal}{\emph{Scientific Reports}} \bibinfo{volume}{14}, \bibinfo{number}{1} (\bibinfo{year}{2024}), \bibinfo{pages}{27446}.
\newblock


\bibitem[Reichman and Heck(2024)]%
        {dpr-4}
\bibfield{author}{\bibinfo{person}{Benjamin Reichman} {and} \bibinfo{person}{Larry Heck}.} \bibinfo{year}{2024}\natexlab{}.
\newblock \showarticletitle{Dense Passage Retrieval: Is it Retrieving?}. In \bibinfo{booktitle}{\emph{Findings of the Association for Computational Linguistics: EMNLP 2024}}. \bibinfo{pages}{13540--13553}.
\newblock


\bibitem[{\c{S}}akar and Emekci(2024)]%
        {ragimp1}
\bibfield{author}{\bibinfo{person}{Tolga {\c{S}}akar} {and} \bibinfo{person}{Hakan Emekci}.} \bibinfo{year}{2024}\natexlab{}.
\newblock \showarticletitle{Maximizing RAG efficiency: A comparative analysis of RAG methods}.
\newblock \bibinfo{journal}{\emph{Natural Language Processing}} (\bibinfo{year}{2024}), \bibinfo{pages}{1--25}.
\newblock


\bibitem[Salemi and Zamani(2024)]%
        {10.1145/3626772.3657957}
\bibfield{author}{\bibinfo{person}{Alireza Salemi} {and} \bibinfo{person}{Hamed Zamani}.} \bibinfo{year}{2024}\natexlab{}.
\newblock \showarticletitle{Evaluating Retrieval Quality in Retrieval-Augmented Generation}. In \bibinfo{booktitle}{\emph{Proceedings of the 47th International ACM SIGIR Conference on Research and Development in Information Retrieval}} (Washington DC, USA) \emph{(\bibinfo{series}{SIGIR '24})}. \bibinfo{publisher}{Association for Computing Machinery}, \bibinfo{address}{New York, NY, USA}, \bibinfo{pages}{2395–2400}.
\newblock
\showISBNx{9798400704314}
\urldef\tempurl%
\url{https://doi.org/10.1145/3626772.3657957}
\showDOI{\tempurl}


\bibitem[Schulz et~al\mbox{.}(2017)]%
        {frame}
\bibfield{author}{\bibinfo{person}{Hannes Schulz}, \bibinfo{person}{Jeremie Zumer}, \bibinfo{person}{Layla~El Asri}, {and} \bibinfo{person}{Shikhar Sharma}.} \bibinfo{year}{2017}\natexlab{}.
\newblock \showarticletitle{A frame tracking model for memory-enhanced dialogue systems}.
\newblock \bibinfo{journal}{\emph{arXiv preprint arXiv:1706.01690}} (\bibinfo{year}{2017}).
\newblock


\bibitem[Sciavolino et~al\mbox{.}(2021)]%
        {sparse-vs-dense2}
\bibfield{author}{\bibinfo{person}{Christopher Sciavolino}, \bibinfo{person}{Zexuan Zhong}, \bibinfo{person}{Jinhyuk Lee}, {and} \bibinfo{person}{Danqi Chen}.} \bibinfo{year}{2021}\natexlab{}.
\newblock \showarticletitle{Simple entity-centric questions challenge dense retrievers}.
\newblock \bibinfo{journal}{\emph{arXiv preprint arXiv:2109.08535}} (\bibinfo{year}{2021}).
\newblock


\bibitem[Siriwardhana et~al\mbox{.}(2023)]%
        {ragimp3}
\bibfield{author}{\bibinfo{person}{Shamane Siriwardhana}, \bibinfo{person}{Rivindu Weerasekera}, \bibinfo{person}{Elliott Wen}, \bibinfo{person}{Tharindu Kaluarachchi}, \bibinfo{person}{Rajib Rana}, {and} \bibinfo{person}{Suranga Nanayakkara}.} \bibinfo{year}{2023}\natexlab{}.
\newblock \showarticletitle{Improving the domain adaptation of retrieval augmented generation (RAG) models for open domain question answering}.
\newblock \bibinfo{journal}{\emph{Transactions of the Association for Computational Linguistics}}  \bibinfo{volume}{11} (\bibinfo{year}{2023}), \bibinfo{pages}{1--17}.
\newblock


\bibitem[Song and Zheng(2024)]%
        {query-formulation-6}
\bibfield{author}{\bibinfo{person}{Mingyang Song} {and} \bibinfo{person}{Mao Zheng}.} \bibinfo{year}{2024}\natexlab{}.
\newblock \showarticletitle{A Survey of Query Optimization in Large Language Models}.
\newblock \bibinfo{journal}{\emph{arXiv preprint arXiv:2412.17558}} (\bibinfo{year}{2024}).
\newblock


\bibitem[Tang and Yang(2024)]%
        {multihoprag-paper}
\bibfield{author}{\bibinfo{person}{Yixuan Tang} {and} \bibinfo{person}{Yi Yang}.} \bibinfo{year}{2024}\natexlab{}.
\newblock \showarticletitle{Multihop-rag: Benchmarking retrieval-augmented generation for multi-hop queries}.
\newblock \bibinfo{journal}{\emph{arXiv preprint arXiv:2401.15391}} (\bibinfo{year}{2024}).
\newblock


\bibitem[Wang et~al\mbox{.}(2023)]%
        {ragf}
\bibfield{author}{\bibinfo{person}{Liang Wang}, \bibinfo{person}{Nan Yang}, {and} \bibinfo{person}{Furu Wei}.} \bibinfo{year}{2023}\natexlab{}.
\newblock \showarticletitle{Query2doc: Query expansion with large language models}.
\newblock \bibinfo{journal}{\emph{arXiv preprint arXiv:2303.07678}} (\bibinfo{year}{2023}).
\newblock


\bibitem[Wang et~al\mbox{.}(2024)]%
        {rag2}
\bibfield{author}{\bibinfo{person}{Xiaohua Wang}, \bibinfo{person}{Zhenghua Wang}, \bibinfo{person}{Xuan Gao}, \bibinfo{person}{Feiran Zhang}, \bibinfo{person}{Yixin Wu}, \bibinfo{person}{Zhibo Xu}, \bibinfo{person}{Tianyuan Shi}, \bibinfo{person}{Zhengyuan Wang}, \bibinfo{person}{Shizheng Li}, \bibinfo{person}{Qi Qian}, {et~al\mbox{.}}} \bibinfo{year}{2024}\natexlab{}.
\newblock \showarticletitle{Searching for best practices in retrieval-augmented generation}. In \bibinfo{booktitle}{\emph{Proceedings of the 2024 Conference on Empirical Methods in Natural Language Processing}}. \bibinfo{pages}{17716--17736}.
\newblock


\bibitem[Wu et~al\mbox{.}(2024)]%
        {rag1}
\bibfield{author}{\bibinfo{person}{Shangyu Wu}, \bibinfo{person}{Ying Xiong}, \bibinfo{person}{Yufei Cui}, \bibinfo{person}{Haolun Wu}, \bibinfo{person}{Can Chen}, \bibinfo{person}{Ye Yuan}, \bibinfo{person}{Lianming Huang}, \bibinfo{person}{Xue Liu}, \bibinfo{person}{Tei-Wei Kuo}, \bibinfo{person}{Nan Guan}, {et~al\mbox{.}}} \bibinfo{year}{2024}\natexlab{}.
\newblock \showarticletitle{Retrieval-augmented generation for natural language processing: A survey}.
\newblock \bibinfo{journal}{\emph{arXiv preprint arXiv:2407.13193}} (\bibinfo{year}{2024}).
\newblock


\bibitem[Xiong et~al\mbox{.}(2020)]%
        {anncomp1}
\bibfield{author}{\bibinfo{person}{Lee Xiong}, \bibinfo{person}{Chenyan Xiong}, \bibinfo{person}{Ye Li}, \bibinfo{person}{Kwok-Fung Tang}, \bibinfo{person}{Jialin Liu}, \bibinfo{person}{Paul Bennett}, \bibinfo{person}{Junaid Ahmed}, {and} \bibinfo{person}{Arnold Overwijk}.} \bibinfo{year}{2020}\natexlab{}.
\newblock \showarticletitle{Approximate nearest neighbor negative contrastive learning for dense text retrieval}.
\newblock \bibinfo{journal}{\emph{arXiv preprint arXiv:2007.00808}} (\bibinfo{year}{2020}).
\newblock


\bibitem[Yang et~al\mbox{.}(2018)]%
        {hotpotqa}
\bibfield{author}{\bibinfo{person}{Zhilin Yang}, \bibinfo{person}{Peng Qi}, \bibinfo{person}{Saizheng Zhang}, \bibinfo{person}{Yoshua Bengio}, \bibinfo{person}{William~W Cohen}, \bibinfo{person}{Ruslan Salakhutdinov}, {and} \bibinfo{person}{Christopher~D Manning}.} \bibinfo{year}{2018}\natexlab{}.
\newblock \showarticletitle{HotpotQA: A dataset for diverse, explainable multi-hop question answering}.
\newblock \bibinfo{journal}{\emph{arXiv preprint arXiv:1809.09600}} (\bibinfo{year}{2018}).
\newblock


\bibitem[Zamani et~al\mbox{.}(2022)]%
        {ragg}
\bibfield{author}{\bibinfo{person}{Hamed Zamani}, \bibinfo{person}{Michael Bendersky}, \bibinfo{person}{Donald Metzler}, \bibinfo{person}{Honglei Zhuang}, {and} \bibinfo{person}{Xuanhui Wang}.} \bibinfo{year}{2022}\natexlab{}.
\newblock \showarticletitle{Stochastic retrieval-conditioned reranking}. In \bibinfo{booktitle}{\emph{Proceedings of the 2022 ACM SIGIR International Conference on Theory of Information Retrieval}}. \bibinfo{pages}{81--91}.
\newblock


\bibitem[Zhao et~al\mbox{.}(2024)]%
        {rag3}
\bibfield{author}{\bibinfo{person}{Penghao Zhao}, \bibinfo{person}{Hailin Zhang}, \bibinfo{person}{Qinhan Yu}, \bibinfo{person}{Zhengren Wang}, \bibinfo{person}{Yunteng Geng}, \bibinfo{person}{Fangcheng Fu}, \bibinfo{person}{Ling Yang}, \bibinfo{person}{Wentao Zhang}, {and} \bibinfo{person}{Bin Cui}.} \bibinfo{year}{2024}\natexlab{}.
\newblock \showarticletitle{Retrieval-augmented generation for ai-generated content: A survey}.
\newblock \bibinfo{journal}{\emph{arXiv preprint arXiv:2402.19473}} (\bibinfo{year}{2024}).
\newblock


\bibitem[Zhuang et~al\mbox{.}(2023)]%
        {qgen2}
\bibfield{author}{\bibinfo{person}{Yuchen Zhuang}, \bibinfo{person}{Yue Yu}, \bibinfo{person}{Kuan Wang}, \bibinfo{person}{Haotian Sun}, {and} \bibinfo{person}{Chao Zhang}.} \bibinfo{year}{2023}\natexlab{}.
\newblock \showarticletitle{Toolqa: A dataset for llm question answering with external tools}.
\newblock \bibinfo{journal}{\emph{Advances in Neural Information Processing Systems}}  \bibinfo{volume}{36} (\bibinfo{year}{2023}), \bibinfo{pages}{50117--50143}.
\newblock


\end{thebibliography}



\end{document}